\documentclass[%
 aip,
 rsi,
 amsmath,amssymb,
 reprint,%
]{revtex4-1}

\usepackage{graphicx}
\usepackage{dcolumn}
\usepackage{bm}

\usepackage[utf8]{inputenc}
\usepackage[T1]{fontenc}
\usepackage{mathptmx}

\usepackage{xcolor}

\begin{document}

\title{
High-purity solid parahydrogen
}

\author{Ashok Bhandari}
\affiliation{Department of Physics, University of Nevada, Reno NV 89557, USA}
\author{Alexandar P. Rollings}
\affiliation{Department of Physics, University of Nevada, Reno NV 89557, USA}
\author{Levi  Ratto}
\affiliation{Department of Physics, University of Nevada, Reno NV 89557, USA}
\author{Jonathan D. Weinstein}
\email{weinstein@physics.unr.edu}
\homepage{http://www.weinsteinlab.org}
\affiliation{Department of Physics, University of Nevada, Reno NV 89557, USA}


\begin{abstract}
Alkali atoms trapped in solid hydrogen matrices have demonstrated ultralong electron spin coherence times, and are promising as quantum sensors. Their spin coherence is limited by magnetic noise from naturally-occurring orthohydrogen molecules in the parahydrogen matrix. 
In the gas phase, the orthohydrogen component of hydrogen can be converted to parahydrogen by flowing it over a catalyst held at cryogenic temperatures, with lower temperatures giving a lower orthohydrogen fraction.
In this work, we use a single cryostat to reduce the orthohydrogen fraction of hydrogen gas and grow a solid matrix from the resulting high-purity parahydrogen. We demonstrate operation of the catalyst down to a temperature of 8~K, and we spectroscopically verify that orthohydrogen impurities in the resulting solid are at a level $< 10^{-6}$. We also find that, at sufficiently low temperatures, the cryogenic catalyst provides isotopic purification, reducing the HD fraction.
\end{abstract}

\maketitle

\section{Introduction}

Solid hydrogen is a molecular solid of H$_2$ \cite{RevModPhys.52.393}. The $^{1}\Sigma$ ground state of the H$_2$ molecule has no net electronic angular momentum or magnetic moment. However, the molecule exists in two possible nuclear spin states: $I=0$ parahydrogen and $I=1$ orthohydrogen. Parahydrogen is nonmagnetic, while orthohydrogen has a small magnetic moment from its nuclear spin. 

Parahydrogen has been shown to be an excellent host matrix for ``matrix isolation'' experiments which trap atoms and molecules within a weakly-bound inert matrix.
Because their   interaction with the host matrix are weak, implanted atoms and molecules retain much of their gas-phase properties.
Parahydrogen matrix isolation experiments have  traditionally been used to perform molecular spectroscopy; very narrow lines have been observed in infrared spectroscopy \cite{Momose1998}.

Recent experiments have shown that atoms trapped in solid hydrogen also retain their key properties for use as quantum sensors for magnetic fields: it is possible to control and measure the spin states of the implanted atoms through optical techniques
 \cite{upadhyay2016longitudinal, PhysRevA.100.063419}, and the trapped atoms exhibit both long ensemble spin dephasing times ($T_2^*$) \cite{PhysRevA.100.063419, PhysRevB.100.024106} and long spin coherence times ($T_2$) \cite{PhysRevLett.125.043601}.
However, the coherence time $T_2$ of the electron spin states of the implanted atoms was found to be limited by orthohydrogen impurities in the solid  \cite{PhysRevLett.125.043601}.
In separate experiments, NMR measurements of HD molecules in solid parahydrogen showed that the ensemble nuclear spin dephasing time $T_2^*$ of the HD molecules was also limited by orthohydrogen \cite{delrieu1981quantum, washburn1981nmr}.
This is similar to the behavior observed for other solid-state quantum sensors, such as NV centers in diamond, in which NV coherence times are limited by the ``nuclear spin bath'' of the $^{13}$C nuclei in the diamond \cite{childress2006coherent}. For NV centers this limitation has been addressed with the development of isotopically purified diamond samples \cite{barry2020sensitivity, teraji2013effective}.
For future work seeking to use atoms and molecules in solid parahydrogen as quantum sensors, it is crucial to produce low-orthohydrogen samples of parahydrogen.


In the absence of a catalyst, the conversion of a hydrogen molecule between the orthohydrogen and parahydrogen states is extremely slow, even in the solid phase \cite{RevModPhys.52.393, schmidt1974diffusion, shevtsov2000quantum}; for the orthohydrogen fractions explored in the current work, there is negligible conversion on the timescale of days.

Conversion between ortho- and para- states can be sped through the use of a paramagnetic catalyst; in the current work we use iron oxide, as detailed in section \ref{sec:apparatus}. If exposed to the catalyst for a sufficient amount of time, the ortho- and para- populations should reach thermal equilibrium. As parahydrogen is the lower-energy state, in the $T\rightarrow0$ limit, the fraction of molecules in the orthohydrogen state should go to zero  \cite{RevModPhys.52.393}.

Prior work has explored the use of a variety of catalysts and methods of sample growth   \cite{oka1993high, tam1999ortho, wu2004infrared, andrews2004simple, tom2009producing, sundararajan2016production, tsuge2020spectroscopy}. 
The lowest orthohydrogen fractions are likely obtained by implanting high densities of catalyst atoms or molecules directly into solid parahydrogen, where they can serve to achieve para-ortho thermal equilibrium at arbitrarily low temperatures \cite{PhysRevLett.125.043601, yoshioka2003infrared}.
However, because the catalyst itself has undesirable magnetic properties, in our work the parahydrogen must be extracted from the catalyst for deposition elsewhere.  Some prior work implied it was impractical to extract hydrogen if the catalyst temperature is much below the triple point of hydrogen at 13.8~K \cite{yoshioka2006infrared, fajardo2011matrix}, which would limit the orthohydrogen fraction to levels $\gtrsim 10^{-5}$.
In the prior literature, catalyst temperatures down to $\sim 13$~K were reported, and orthohydrogen fractions $\lesssim 10^{-4}$ were verified spectroscopically \cite{ tam1999ortho,yoshioka2003infrared, tom2009producing, tsuge2020spectroscopy}.

In the current work, we operate at temperatures both above and below the liquid-solid phase transition of hydrogen and extract the hydrogen from the catalyst as a gas; the vapor pressure of hydrogen (shown below in Fig. \ref{fig:VaporPressures}), limits how cold the catalyst can be. We show that we can operate at temperatures as low as 8~K --- a temperature at which the equilibrium orthohydrogen fraction is over three orders of magnitude lower than the triple point
--- and spectroscopically verify orthohydrogen fractions $ < 10^{-6}$.

\section{Apparatus}
\label{sec:apparatus}

\begin{figure}[ht]
    \begin{center}
    \includegraphics[width=\linewidth]{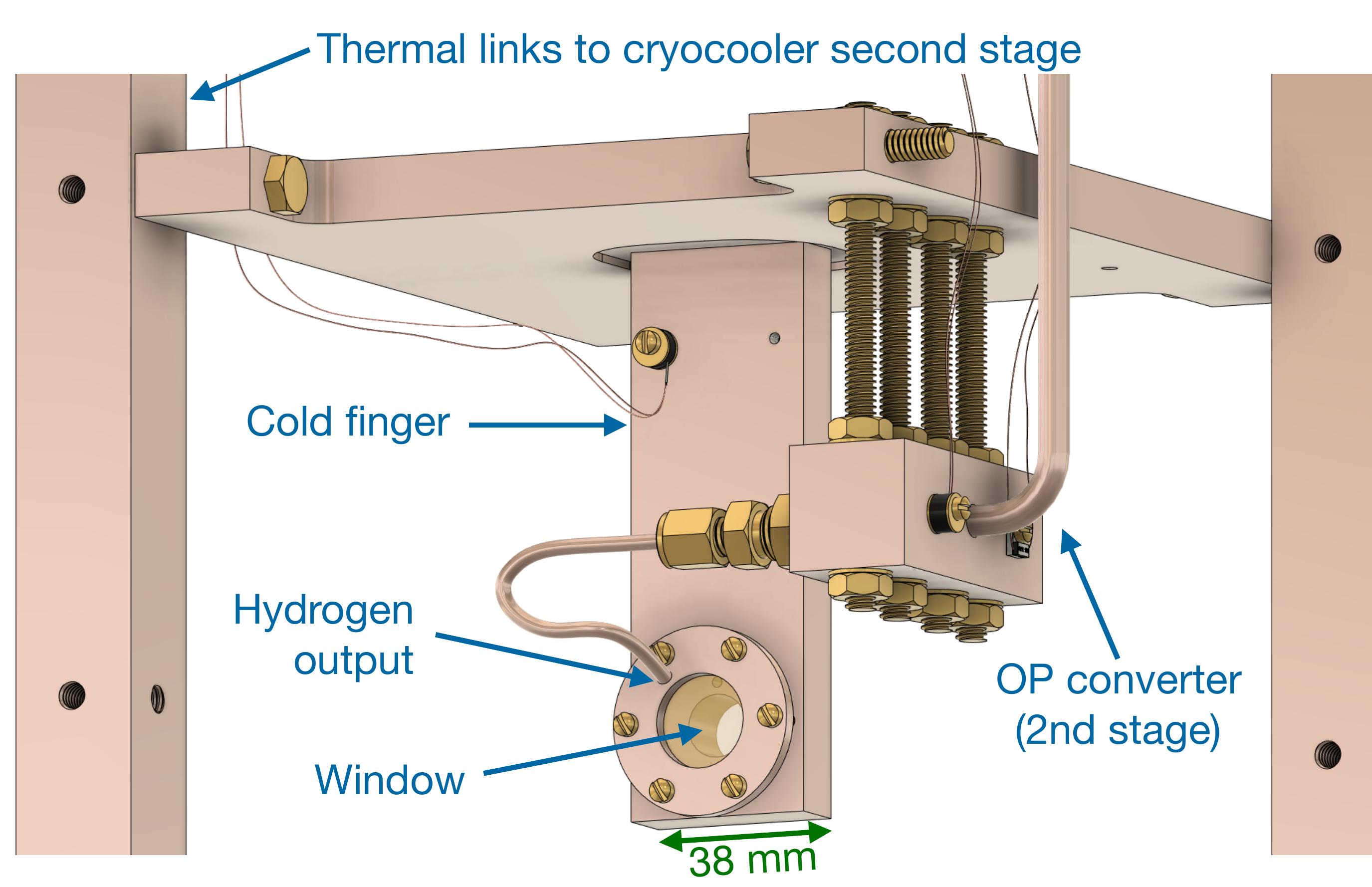}
    \caption{ 
Schematic of the ``heart'' of the apparatus. All parts shown are copper, with the exception of the brass hardware, the sapphire window, the thermometer mounted on the coldfinger, and the thermometer and heater mounted on the OP converter. 
The structural components are copper alloys 101 and 110; the copper tubing is alloy 122. The thermometers and heaters are depicted on the faces opposite their actual location for visibility. There are four rigid thermal links to the cryocooler second stage; two are removed for clarity.
 \label{fig:apparatusHeart}
    }
    \end{center}
\end{figure}
\begin{figure}[ht]
    \begin{center}
    \includegraphics[width=0.8\linewidth]{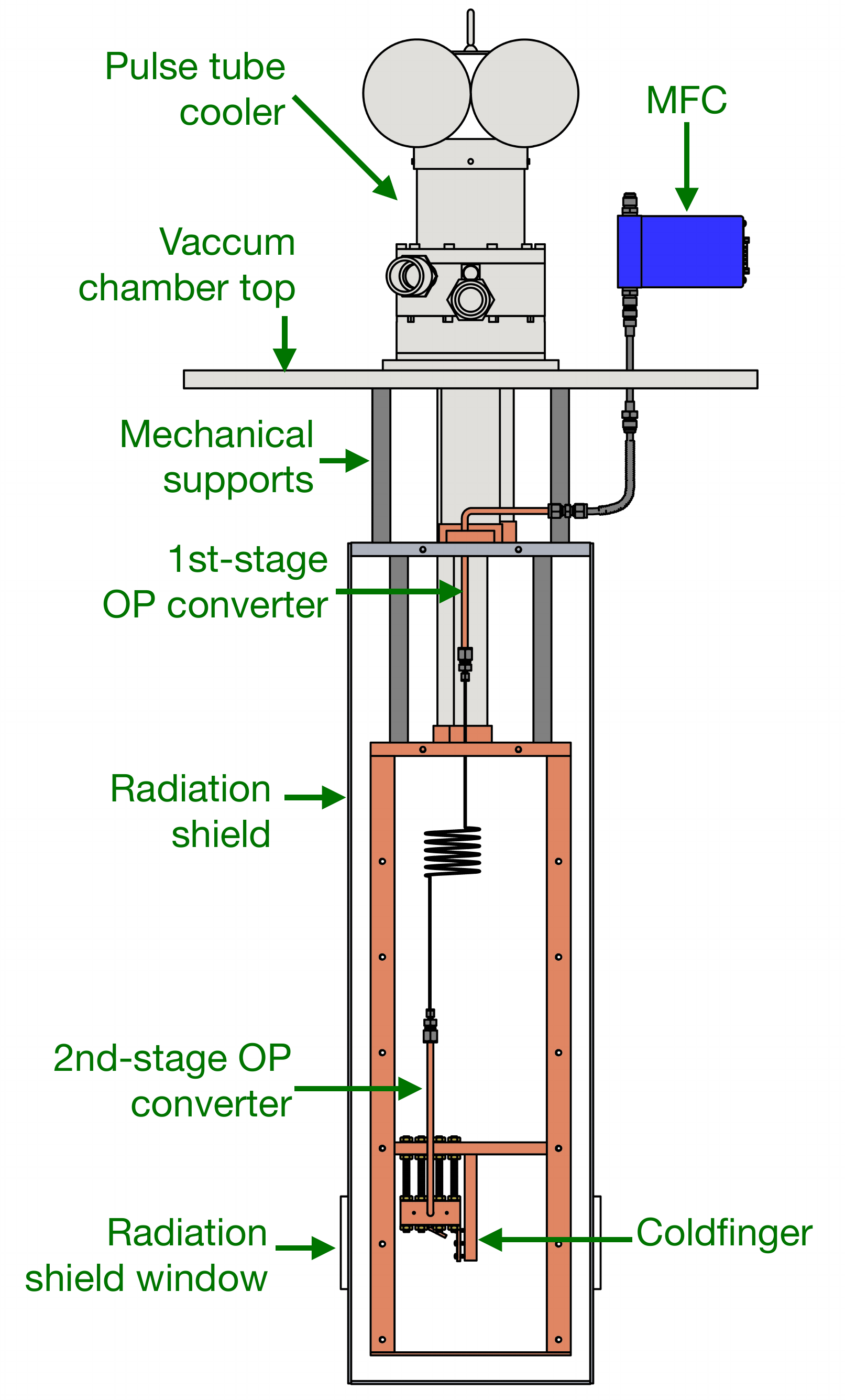}
    \caption{ 
Schematic of the apparatus, as discussed in the text. The vacuum chamber is omitted for clarity, and various elements and vacuum feedthroughs have similarly been simplified or omitted.
The FTIR beam is transmitted through the sample through the radiation shield windows.
 \label{fig:apparatus}
    }
    \end{center}
\end{figure}

Schematics of our apparatus are shown in figures \ref{fig:apparatusHeart} and \ref{fig:apparatus}.

Our samples are grown by vacuum deposition of hydrogen gas onto a antireflection-coated sapphire window. The window is thermally connected to a copper coldfinger via a thin layer of indium. The coldfinger is cooled by the second stage of a pulse-tube cooler through a series of rigid and flexible copper heatlinks. The cold finger base temperature is 3.6~K, as measured by a silicon diode thermometer. Unless otherwise stated, all temperatures reported in this paper have an accuracy of $\pm 0.5$~K, limited by the calibration accuracy of our silicon diode thermometers.

To control the orthohydrogen fraction, we flow the hydrogen through two ``OP converters'' (ortho-para converters) prior to deposition.
Each OP converter is made using 0.25" outer-diameter, 0.19" inner-diameter ``refrigeration tubing'' which contains our ortho-para catalyst.
Before being filled with catalyst, each copper tube is soldered into a 1/4"-diameter through-hole in a copper block for a thermal and mechanical connection, as shown in Figs. \ref{fig:apparatusHeart} and \ref{fig:apparatus}. 
The through-hole in the copper block is 13~mm long in the case of the first-stage OP converter and 25~mm long in the case of the second-stage OP converter.
The catalyst used is iron (III) oxide in powder form (30-50 mesh) \cite{Sigma_catalyst}.
The ``first-stage'' and ``second-stage'' OP converters consist of 40~cm and 30~cm lengths of tubing, filled with 9 and 7 grams of catalyst, respectively. To keep the granular catalyst inside the tubing, glass wool is inserted in each end of each tube; the glass wool is held in place by the Swagelok-style compression fittings used to connect the different sections of our hydrogen gas ``plumbing''. 

The first-stage OP converter is thermally anchored to the first stage of our cryocooler, as shown in figure \ref{fig:apparatus}. During hydrogen flow, its temperature is $42.5 \pm 2.5$~K. The second-stage OP converter is thermally anchored to the second stage of our cryocooler through a heat link consisting of 1 to 4 brass 1/4"-20 threaded rods, 0.1~m in length, as shown in Fig. \ref{fig:apparatus}.  The temperature of the second-stage OP converter is controlled by a resistive heater and measured with a silicon diode thermometer mounted to the OP converter's copper block. The weak heatlink allows us to maintain the second-stage OP converter at an elevated temperature without an undue thermal load on the pulse-tube refrigerator; the conductivity of the heatlink is adjusted by varying the number of threaded rods.
During deposition we can vary the temperature of the second OP converter from 7.5~K to 30~K while maintaining cold finger temperatures $\lesssim 4.5$~K.

The flow of hydrogen gas is controlled by a room-temperature mass-flow controller. 
The connections from the mass-flow-controller to the first-stage OP converter and from the first-stage to the second-stage OP converter are made with thin-walled stainless steel tubing for thermal isolation.
After exiting the second-stage OP converter, the hydrogen gas flows through a short length of 1/8" diameter copper tubing, aimed at the sapphire window, as seen in Fig. \ref{fig:apparatusHeart}. The end of the tube is roughly 3~cm from the window surface. 

The coldfinger, the second-stage OP converter, and their thermal links to the second stage of the pulse-tube cooler are surrounded by a radiation shield thermally connected to the first stage of the pulse-tube cooler, as shown in figure \ref{fig:apparatus}. All this is contained within a vacuum chamber formed from ISO~400 nipples. The vacuum chamber, radiation shield, and the copper heatlinks between the pulse-tube cooler and the cold finger were all inherited from a prior experiment \cite{PhysRevLett.108.203201}: we believe the same experimental results could be obtained in a much more compact apparatus. For example, the only reason for the large vertical separation between the coldfinger and the bottom of the pulse tube cooler is so that our sample would be aligned with preexisting windows in the vacuum chamber.

\section{Sample growth conditions}

Unless otherwise stated, all samples in this paper were grown with a hydrogen flow rate of 4~sccm, as controlled by the mass-flow controller. During sample growth, we pump on our vacuum chamber with a small turbomolecular pump with a nominal 80~L/s pumping speed. While the majority of our pumping is cryopumping from the cold surfaces in our chamber, the turbopump is important to evacuate the small amount of helium gas present in our hydrogen gas. As measured by a residual gas analyzer, the partial pressures of hydrogen and helium during sample growth are $10^{-6}$ and $10^{-9}$~Torr, respectively, increased from their background pressures of $10^{-9}$ and $\lesssim 10^{-10}$~Torr when we do not flow hydrogen gas.

During growth, the sample thickness is monitored through infrared spectroscopy. Light from an FTIR spectrometer is sent through our sample and onto an ``external'' mercury cadmium telluride detector; the resolution of the spectrometer was 0.125~cm$^{-1}$. The beam size at the sample is 4~mm by 5~mm (FWHM). This same spectroscopic technique was used after sample growth was completed to measure the orthohydrogen fraction of the sample, as discussed in section \ref{sec:DataAnalysis}.

Sample thicknesses ranged from 0.4 to 4~mm; 
deposition rates are discussed in section \ref{sec:DepositionRates}. Samples were grown for second-stage OP converter temperatures ranging from 8.1~K to 29~K.
An attempt to grow a sample at an OP converter temperature of 7.6~K initially exhibited abnormally slow growth, followed by a stoppage of hydrogen flow (as measured by the mass-flow controller). We attribute this to clogging of the OP converter due to frozen hydrogen: at this temperature, the vapor pressure of H$_2$ (see Fig. \ref{fig:VaporPressures}) is insufficient to avoid condensation at a 4~sccm flow rate.

During sample growth, the cold finger temperature ranged from 3.8~K to 4.6~K, up from its base temperature due to the heat load of the depositing gas and the heater used to maintain the OP converter at elevated temperatures.

\section{Spectroscopy}
\label{sec:DataAnalysis}

\begin{figure}[ht]
    \begin{center}
    \includegraphics[width=\linewidth]{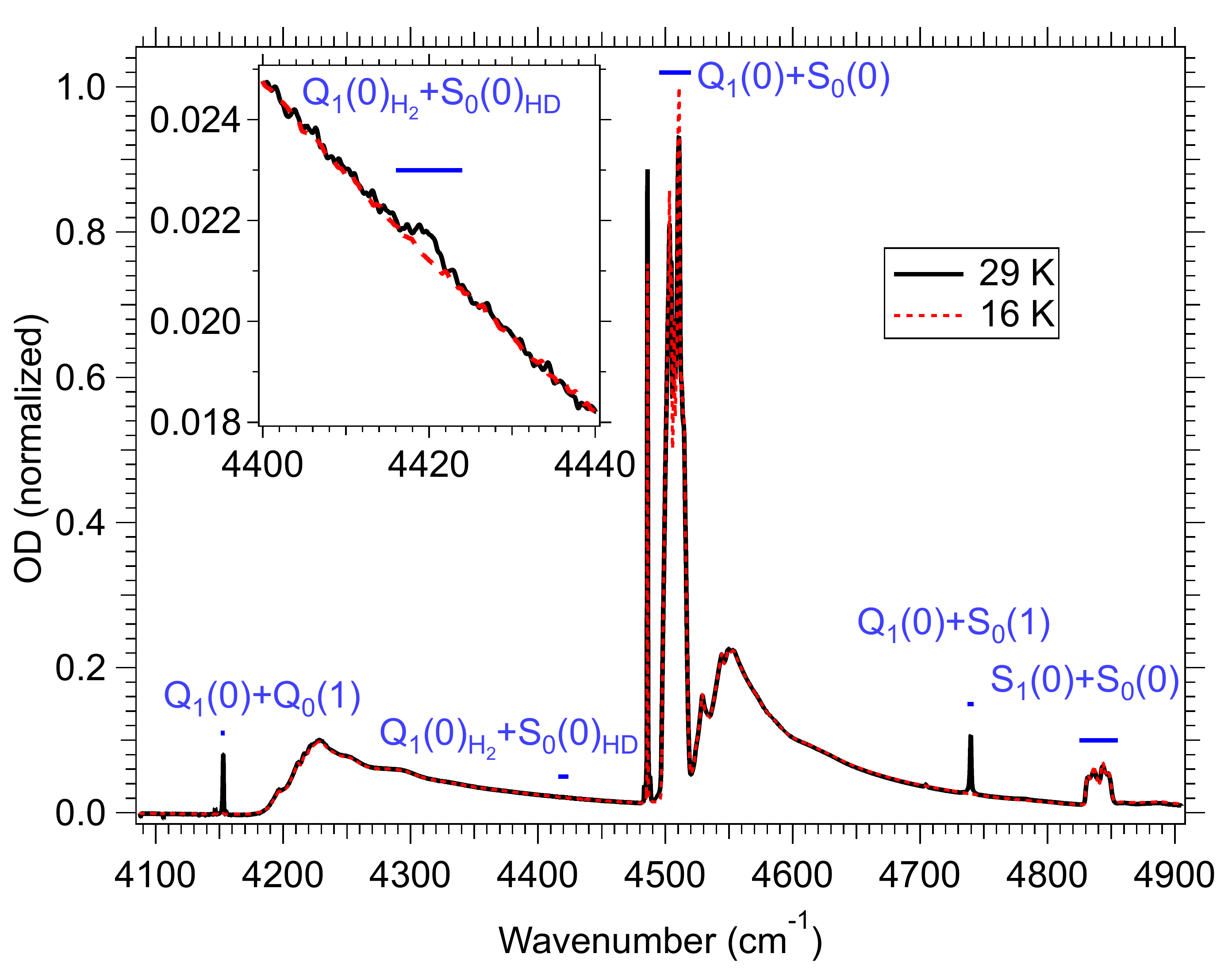}
    \caption{ 
Spectra of two samples grown at OP converter temperatures of 16 and 29~K.
To simplify visual comparison of the two spectra, the background OD has been subtracted, the spectra have been normalized by the area of the $Q_1(0) + S_0(0)$ transition, and the spectra have been low-pass filtered to reduce the resolution to $\sim 0.5$~cm$^{-1}$. These operations are not done on the spectra used for analysis of our samples.
The spectral features used in this paper are labeled according to the notation of references \cite{fajardo2019solid, crane1966induced, van1968theory}. 
The inset shows the same spectrum as the main figure, ``zoomed in'' to show the transition used to measure HD molecules.
\label{fig:IRspectra}
    }
    \end{center}
\end{figure}

Sample spectra of two parahydrogen samples are shown in Fig. \ref{fig:IRspectra}. 
We determine the transmission $T$ of the parahydrogen sample by comparing spectra taken with and without the sample present. We convert the transmission to an the optical depth (OD) using the convention that the $T \equiv e^{-\mathrm{OD}}$. 
We filter all our spectra to remove interference fringes from etalon effects from the multiple cryostat windows that the beam passes through. 

We determine the thickness $t$ of the sample using the Q$_1$(0) + S$_0$(0) and S$_1$(0) + S$_0$(0) transitions, following the procedure outlined by Fajardo \cite{fajardo2019solid}.
\begin{equation}
     t = 4.8	\times 10^{-2} \ \mathrm{mm}
     \cdot
     \int_{4495 \ \mathrm{cm}^{-1}}^{4520 \ \mathrm{cm}^{-1}} \mathrm{OD} \,d\mathrm{n} 
     /
     \mathrm{cm}^{-1}
     \label{eq:thick1}
\end{equation}
\begin{equation}
      t = 6.2 \times10^{-1} \ \mathrm{mm}
      \cdot
      \int_{4825\ \mathrm{cm}^{-1}}^{4855 \ \mathrm{cm}^{-1}} \mathrm {OD} \,d\mathrm{n} 
      /
      \mathrm{cm}^{-1}
           \label{eq:thick2}
\end{equation}
In both equations \ref{eq:thick1} and \ref{eq:thick2}, the background OD is subtracted from the integral under the assumption that it is equal to a linear interpolation of the OD at the endpoints of the integral, as per reference \cite{fajardo2019solid}. The estimated thickness error for these formulae is $\pm 3\%$ in the limit of low orthohydrogen fraction \cite{fajardo2019solid}. For our data, we find the two transitions give similar values of $t$, with a standard deviation of 1.3\%.

As seen in Fig. \ref{fig:IRspectra}, the optical depths of the  $\mathrm{Q}_1(0)+\mathrm{Q}_0(1)$ and $\mathrm{Q}_1(0)+\mathrm{S}_0(1)$ ``ortho-induced'' transitions (at 4150 and 4740 cm$^{-1}$, respectviely)
depend on the orthohydrogen fraction in the sample.
The dependence of the $\mathrm{Q}_1(0)+\mathrm{Q}_0(1)$ transition's optical depth on the orthohydrogen fraction $f_\mathrm{ortho}$ has previously been reported in the literature. In the low-ortho limit, Fajardo \cite{fajardo2011matrix} reports that:
\begin{equation}
      f_\mathrm{ortho} = 
      \frac{{1.24} \times10^{-1} \ \mathrm{mm}}{t}
      \cdot
       \int_{4151\ \mathrm{cm}^{-1}}^{4154 \ \mathrm{cm}^{-1}} \mathrm {OD} \,d\mathrm{n} 
       /
       \mathrm{cm}^{-1}
\end{equation}
with the background OD subtracted from the integral under the assumption that the background is a linear interpolation of the OD at the endpoints of integration. 
The accuracy of this formula is reported to be $\pm 10\%$ \cite{fajardo2011matrix}.

\begin{figure}[ht]
    \begin{center}
    \includegraphics[width=\linewidth]{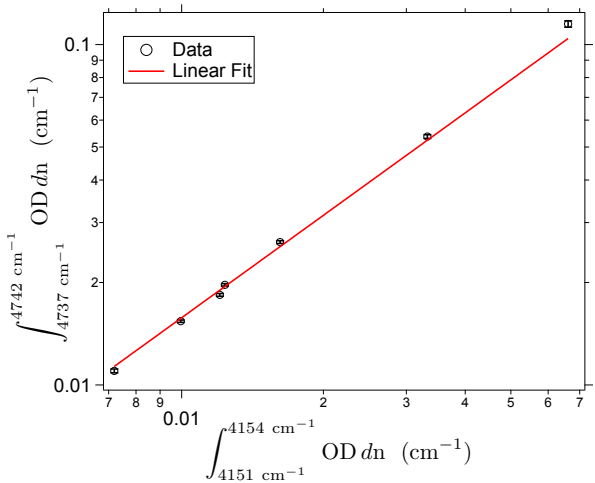}
    \caption{ 
Calibration of the $\mathrm{Q}_1(0)+\mathrm{S}_0(1)$ line (the vertical axis) from the  $\mathrm{Q}_1(0)+\mathrm{Q}_0(1)$ line (the horizontal axis). The data is from samples grown at temperatures from 19~K to 29~K. The background optical depth is subtracted from each integral, as described in the text. \label{fig:OrthoLineCalibration}
    }
    \end{center}
\end{figure}

In this work, we choose to use the $\mathrm{Q}_1(0)+\mathrm{S}_0(1)$ transition to measure the orthohydrogen fraction. While it has the disadvantage of sitting on the broad shoulder of the $\mathrm{S}_1(0)$ phonon sideband,
it is better separated from atmospheric absorption lines and, for our beam path and optics, has better signal-to-noise. We were initially unable to find a literature value for the relation between the $\mathrm{Q}_1(0)+\mathrm{S}_0(1)$ optical depth and the orthohydrogen fraction, so we calibrated it using the 
$\mathrm{Q}_1(0)+\mathrm{Q}_0(1)$ absorption feature, as shown in Fig. \ref{fig:OrthoLineCalibration}.

From this calibration, we find:
\begin{equation}
      f_\mathrm{ortho} = 
      \frac{{7.87} \times10^{-2} \ \mathrm{mm}}{t}
      \cdot
      \int_{4737\ \mathrm{cm}^{-1}}^{4742 \ \mathrm{cm}^{-1}} \mathrm {OD} \,d\mathrm{n}
      /
      \mathrm{cm}^{-1}
      \label{eq:orthofrac}
\end{equation}
The background optical depth is subtracted from the integral as described at the end of this section.

We find there is a small systematic error in formula \ref{eq:orthofrac}, as
the $Q_{1}(0)+S_{0}(1)$ transition was observed to broaden with increasing orthohydrogen fractions.
For OP converter temperatures  between 16 and 29 K, the measured linewidth increases roughly linearly from a FWHM of 0.8 to 1.0 cm$^{-1}$. This leads to a variation in the ``fraction'' of the $Q_{1}(0)+S_{0}(1)$ transition contained within the finite region of integration of equation \ref{eq:orthofrac}. We estimate that, for the orthohydrogen fraction range explored in this work, this leads to errors in $f_\mathrm{ortho}$ of $\lesssim 11\%$. At larger orthohydrogen fractions, this error would likely increase.
Combining this error with the statistical error of our calibration error and the claimed accuracy of reference \cite{fajardo2011matrix}, we estimate the accuracy of formula \ref{eq:orthofrac} to be $\pm 15 \%$ over the range of conditions explored in this work.

During the preparation of this manuscript, we became aware of a prior measurement of the $Q_{1}(0)+S_{0}(1)$  transition by Raston, Kettwich, and Anderson \cite{raston2010infrared}. Their method of spectral analysis --- which integrates over a wider wavelength range than equation \ref{eq:orthofrac} --- is capable of providing accurate results at high ortho fractions. Our narrower range of integration provides lower noise for the measurement of low ortho fractions. The  coefficient of equation \ref{eq:orthofrac} is consistent with the transition strength of reference \cite{raston2010infrared} to within the stated errors of the two works.

To measure the HD fraction in the sample, we use the $\mathrm{Q}_1(0)\mathrm{H}_2+\mathrm{S}_0(0)\mathrm{HD}$ transition at 4420~cm$^{-1}$. While we were unable to unable to find a literature value relating the optical depth to the HD fraction $f_{\mathrm{HD}}$, Crane and Gush \cite{crane1966induced} have published spectra of parahydrogen samples with known HD fractions. From their spectra, we find in the low HD limit,
\begin{equation}
      f_\mathrm{HD} = \frac{{8.96} \times10^{-2} \ \mathrm{mm}}{t} 
      { \int_{4416\ \mathrm{cm}^{-1}}^{4424 \ \mathrm{cm}^{-1}} \mathrm {OD} \,d\mathrm{n}}
      /
      \mathrm{cm}^{-1}
\end{equation}
with an estimated accuracy of $\pm 10\%$. 
The background optical depth is subtracted from the integral as described below.

The $\mathrm{Q}_1(0)+\mathrm{S}_0(1)$ and $\mathrm{Q}_1(0)\mathrm{H}_2+\mathrm{S}_0(0)\mathrm{HD}$ transitions used to measure the orthohydrogen and HD fractions are both located on the phonon sideband of parahydrogen transitions. 
To subtract the background, we use the following procedure.
On both sides of the region of interest (ROI) we integrate the optical depth of 5 adjacent regions of the same width as the ROI. The resulting 10 ``background'' points  are fit as a function of their center wavelengths to a 4th-order polynomial. 
(The ROI is excluded from this fit.) We then use this polynomial to calculate the background in the ROI. We estimate the statistical error of our measurement from the standard deviation of the background points from their polynomial fit.

\section{Results: ortho fraction}

\begin{figure}[ht]
    \begin{center}
    \includegraphics[width=\linewidth]{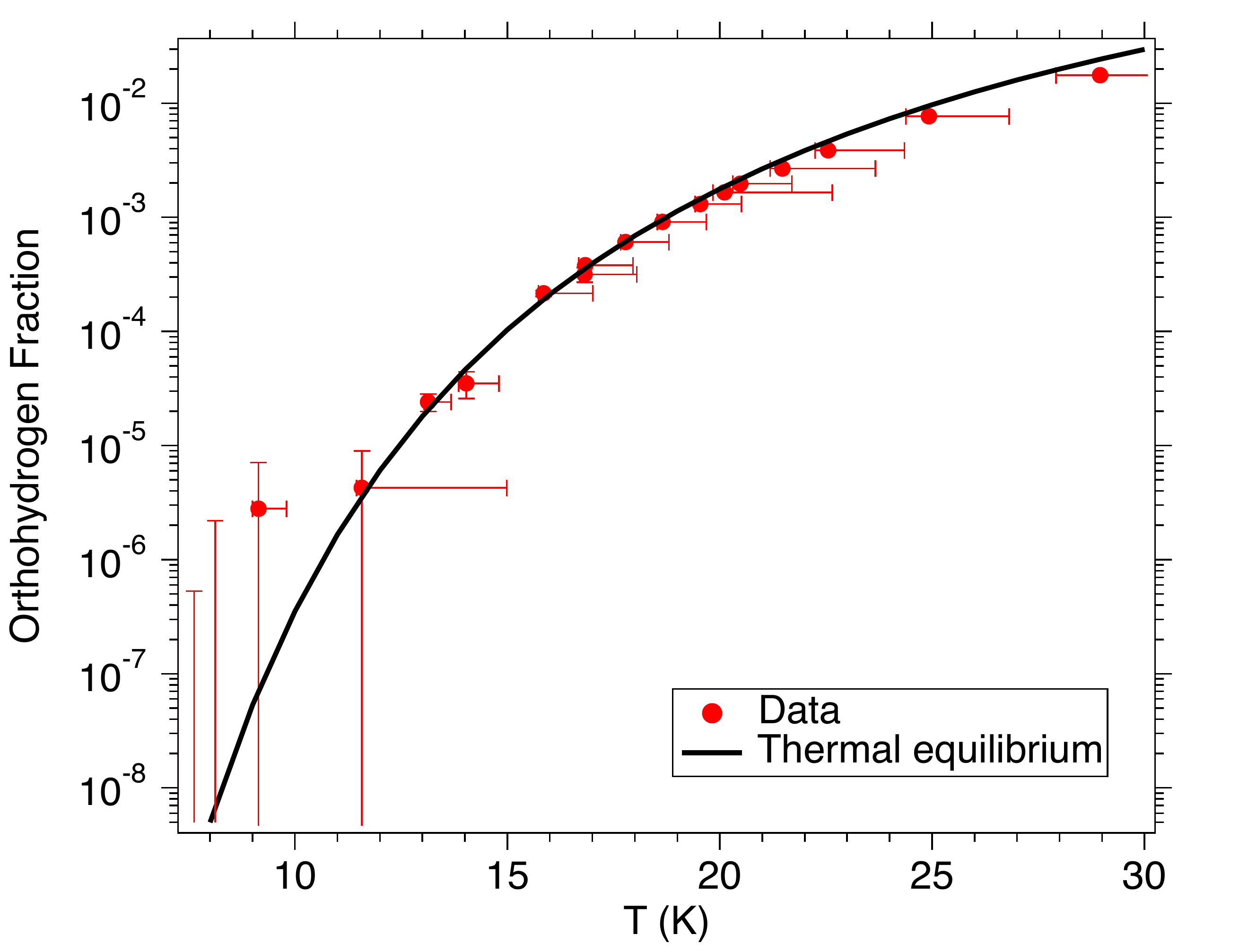}
    \caption{ 
Measurements of the orthohydrogen fraction in deposited samples, plotted as a function of the OP converter temperature during sample growth. The temperature shown is the average temperature of the OP converter during deposition; the temperature error bars indicate the range of temperatures during growth. Not included in the temperature error bar is the $\pm 0.2$~K accuracy of the thermometer. The orthohydrogen fraction error bar represents the statistical error of the meaurement; not included is the $\pm15\%$ uncertainty due to the accuracy of our spectroscopy calibration. For comparison, the expected orthohydrogen fraction in thermal equilibrium is shown as a function of temperature \cite{RevModPhys.52.393}.
\label{fig:OrthoFrac}
    }
    \end{center}
\end{figure}

The results of our measurements of the orthohydrogen fraction of 20 separate samples are shown in 
 Fig. \ref{fig:OrthoFrac}.

We see the expected behavior: to within our experimental error, the measured orthohydrogen fractions are consistent with reaching thermal equilibrium at the temperature of the OP converter. 
At our coldest temperatures, the signal-to-noise of our spectroscopic measurements is less than one, and we can only put an upper limit on the ortho fraction. A weighted average of the data points taken at $T<10$~K gives an upper limit on the ortho fraction of $1\times10^{-6}$ at 95\% confidence. To confirm whether our OP converter is still achieving the expected ``equilibrium'' ortho fractions at temperatures $\lesssim 11$~K will require a more sensitive probe of the ortho fraction than is achievable with our current spectroscopy. If the OP converter is functioning equally well at our coldest temperatures as it was at higher temperatures, the ortho fraction should be $<10^{-8}$.

\section{Results: HD fraction}

Hydrogen gas consists of various isotopic combinations.
There are two naturally-occurring isotopes of atomic hydrogen found on earth: $A=1$ hydrogen and $A=2$ deuterium\cite{NISTAtomicBasic}.
%
Because the ground state of the HD molecule has a nonzero nuclear spin, its magnetic moment may also play a role in limiting the $T_2$ and $T_2^*$ of implanted spin qubits, although this effect has not (to our knowledge) been reported in the literature.

We measure the HD fraction $f_\mathrm{HD}$ spectroscopically as described in section \ref{sec:DataAnalysis}.
At high OP converter temperatures, we see no indication of significant isotopic purification. A weighted average of the results from 7 samples grown at OP converter temperatures $> 19$~K gives $f_\mathrm{HD} = (2.2 \pm 0.6) \times 10^{-4}$; a fraction of $3\times 10^{-4}$ would be expected from the natural abundance of deuterium \cite{NISTAtomicBasic}.
At low OP converter temperatures, the spectroscopic HD signal is undetectably small, and the weighted average of 9 samples grown at OP converter temperatures $<16$~K gives 
an upper limit of  $f_\mathrm{HD} \leq 3 \times 10^{-6}$ (at 95\% confidence). 

\begin{figure}[ht]
    \begin{center}
    \includegraphics[width=\linewidth]{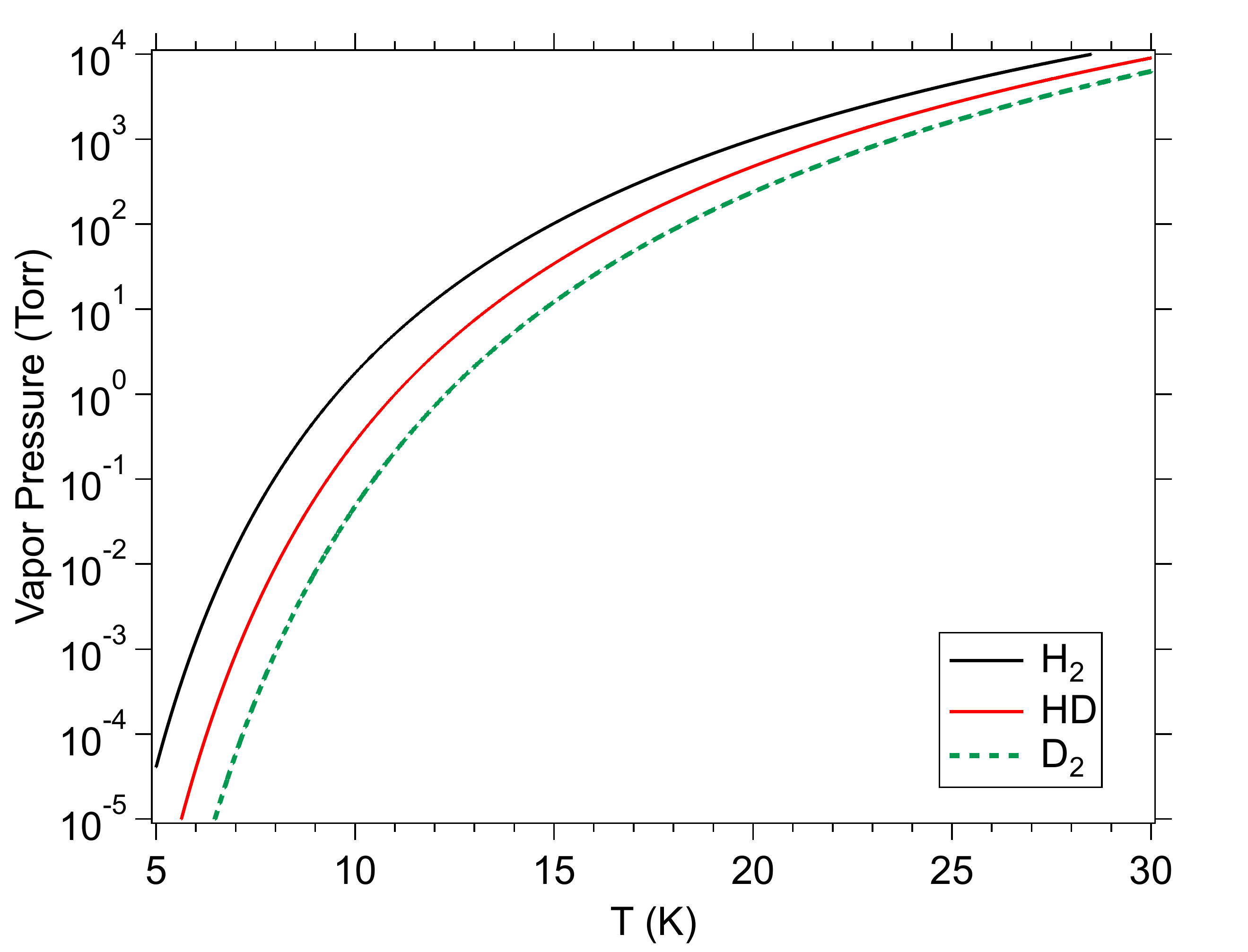}
    \caption{ 
The equilibrium vapor pressures of H$_2$, HD, and D$_2$, from reference \cite{souers1977hydrogen}.
%
%
%
%
%
\label{fig:VaporPressures}
    }
    \end{center}
\end{figure}

While we did not originally anticipate this isotopic purification of the hydrogen gas, in retrospect it is not a surprising result. We speculate the  effect is due to the slight differences in the vapor pressure of the different isotopes of hydrogen, as seen in 
 Fig. \ref{fig:VaporPressures}. These slight differences are likely ``amplified'' by flowing the gas through a tube filled with ortho-para catalyst:
flowing hydrogen gas over a granular media at cryogenic temperatures is a demonstrated method for isotopic purification  \cite{Alekseev2015, junbo2006hydrogen}. We note that prior parahydrogen research noted that HD and D$_2$ condense at higher temperatures than H$_2$, and that this effect can eventually cause clogging of the OP converter
\cite{lorenz2007infrared}. 

\section{Sample growth rate and thermal runaway}
\label{sec:DepositionRates}

\begin{figure}[htb]
    \begin{center}
    \includegraphics[width=\linewidth]{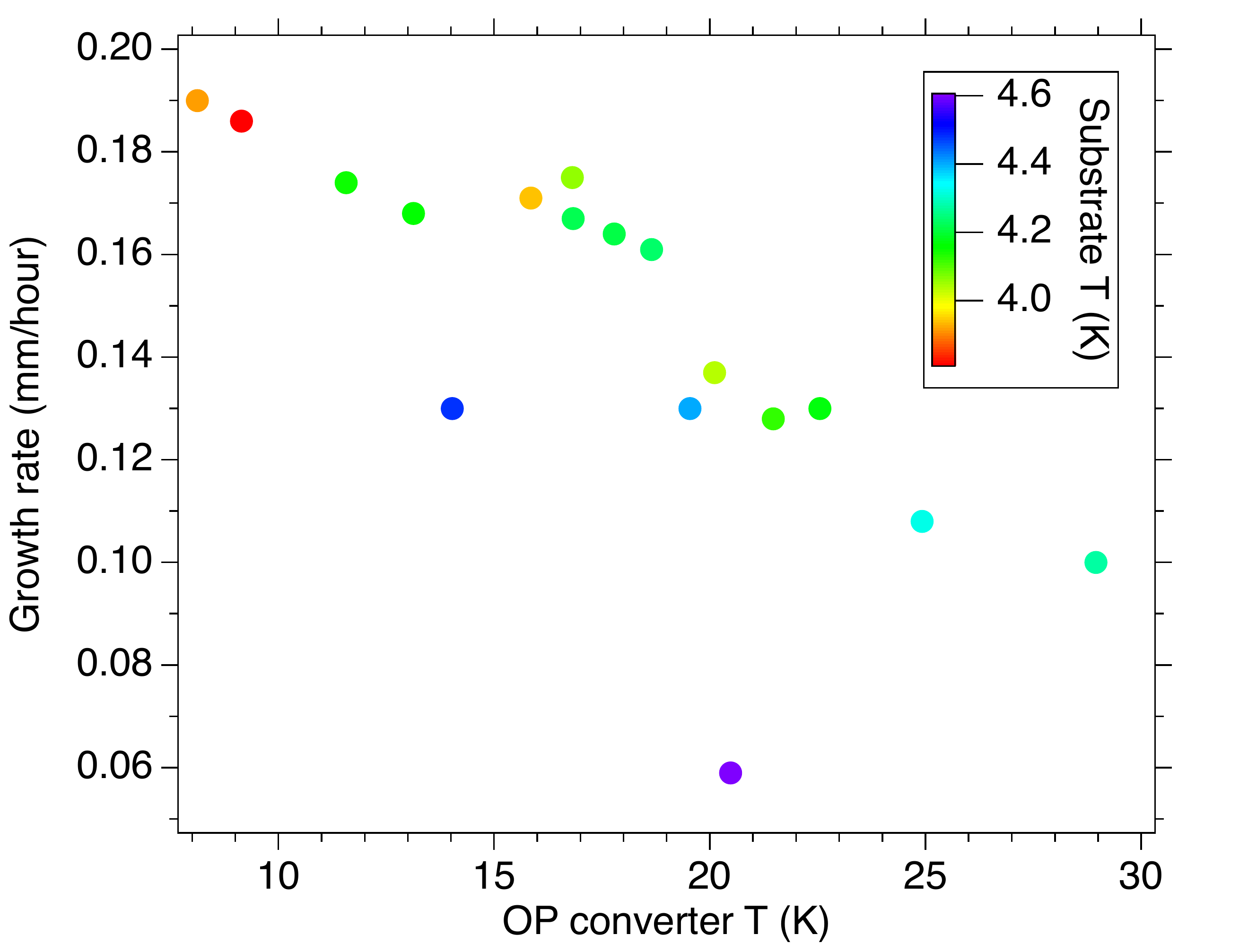}
    \caption{ 
The measured growth rate of different samples, plotted as a function of the average OP converter temperature during sample growth. All samples were grown at a 4~sccm gas flow rate. The data points are colored according to the average substrate temperature during growth. \label{fig:growth_rate}
    }
    \end{center}
\end{figure}

As seen in Fig. \ref{fig:growth_rate}, the growth rate of the sample shows a significant dependence on both the temperature of the second-stage OP converter and, at sufficiently high substrate temperatures, the substrate temperature as well.

The growth rate of the sample is determined by the flux of incoming $H_2$ molecules at the sample surface, their sticking probability, and the sample sublimation rate. 

We speculate that the decrease in the growth rate with increasing OP converter temperature is primarily due to a change in the sticking probability of hydrogen molecules: it has been reported in the literature that the sticking probability of hydrogen molecules (on various surfaces) falls significantly as the temperature of the incident beam increases \cite{matar2010gas,chaabouni2012sticking}.
It is also possible that 
the flux of incoming molecules may --- for a given hydrogen flow rate --- decrease with increasing OP converter temperature due to a possible decrease in the collimation of the incident ``beam'' of molecules. 

As seen from the ``outlying'' data points in Fig. \ref{fig:growth_rate}, it is likely that sublimation of the sample during deposition plays a significiant role in the growth rate at our highest substrate temperatures ($\gtrsim 4.4$~K), but it appears to have little effect on the growth rate at colder substrate temperatures. This is consistent with estimates of sublimation rates using the equilibrium vapor pressure of hydrogen.

We investigated faster sample growth by increasing the hydrogen flow rate. We were able to successfully grow a sample at a flow rate of 17~sccm, which gave a deposition rate of 0.7~mm/hour at an OP converter temperature of 14.8~K and a substrate temperature of 3.7~K.

However, an attempt to grow at a hydrogen flow rate of 21 sccm was unsuccessful, which we attribute to a process of ``thermal runaway''. The heat load of the condensing hydrogen gas causes a rise in the temperature of the coldfinger, which decreases its ability to cryopump the hydrogen gas. This increases the pressure within our cryostat which in turn further increases the coldfinger temperature. At a 21~sccm hydrogen flow, in steady-state we observed a coldfinger temperature of 5.6~K, a cryocooler temperature of 3.5~K (measured at the second-stage of the pulse-tube cooler), and a chamber pressure of $10^{-5}$~Torr. We observed no evidence of sample growth on the coldfinger window.

Other cryocooler-based experiments have grown parahydrogen samples using vacuum deposition at rates of up to 5~mm/hour \cite{tam:1926}. These experiments used two cryocoolers: one to cool the OP converter and one to cool the coldfinger \cite{tam:1926}. 
We speculate that with an improved thermal link between the coldfinger and refrigerator in our apparatus, our single cryocoller design would be capable of sample growth at faster hydrogen flow rates than achieved here.

We note that the sample grown at 17~sccm flow rate exhibited a slightly higher orthohydrogen fraction than expected from thermal equilibrium at the measured OP converter temperature (higher by a factor of 1.3, equivalent to an increase in temperature of 0.4~K). We suspect this is due either to a reduced interaction time with the catalyst or due to heating of the catalyst by the gas flow (as discussed in section \ref{sec:apparatus}, our thermometry does not measure the temperature of the catalyst directly). Reaching low orthohydrogen fractions at significantly higher flow rates than explored in this work may require redesigning the OP converter.

\section{Delay in deposition}

\begin{figure}[ht]
    \begin{center}
    \includegraphics[width=\linewidth]{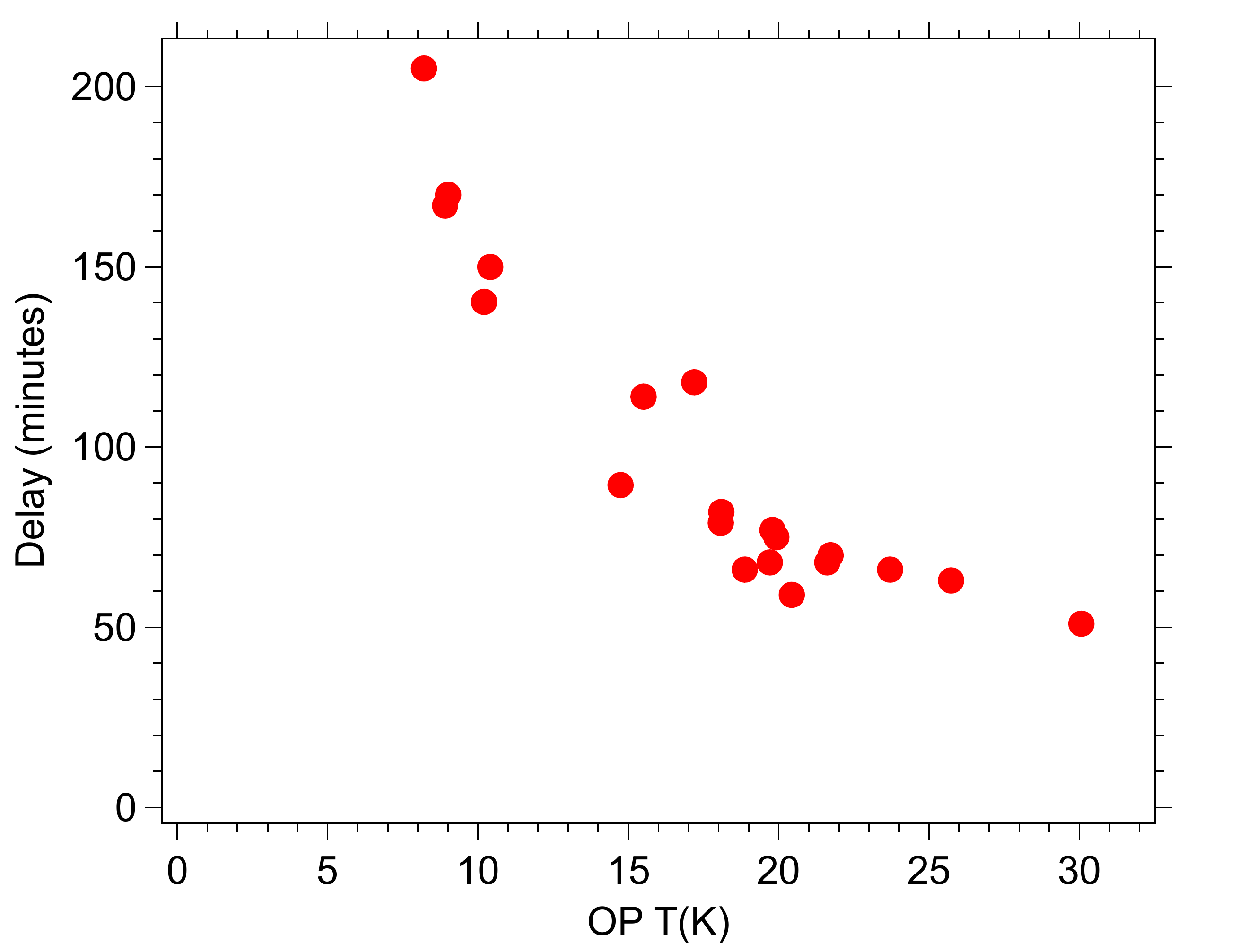}
    \caption{The delay time between the start of hydrogen gas flow and the beginning of sample growth, plotted as a function of the average temperature of the second-stage OP converter during the delay. 
\label{fig:Monolayers}
    }
    \end{center}
\end{figure}

In our sample growth experiments, we observe a significant delay between the start of the hydrogen flow at the room-temperature mass-flow controller and the beginning of sample growth. The delay time is shown in Fig. \ref{fig:Monolayers} for samples grown at a 4~sccm flow rate. A sample grown at a 17~sccm flow rate showed a delay time roughly 4 times shorter than comparable 4~sccm data. In both cases, we observe the heat load on the OP converter from the gas is greater during the delay than it is after deposition begins.

We attribute the delay to the adsorption of hydrogen molecules onto the surface of the catalyst. Once this layer reaches a ``saturation'' thickness, hydrogen exits the OP converter.
We attribute the extra heat load during deposition to the sum of the heat load of cooling the incoming gas (which is also present during steady-state flow) and the heat of deposition (which goes to zero during steady-state flow).

We estimate the thickness of the deposited layer from the delay, prior surface area measurements of the catalyst \cite{hartl2016hydrogen}, and by approximating the density of a thin film of hydrogen to be the same as the bulk density \cite{RevModPhys.52.393}.
From this model, and the assumption that the hydrogen is deposited primarily in the second-stage OP converter, the observed delay at our coldest OP temperatures corresponds to a deposited hydrogen film $2$~monolayers thick, and the delay at a 30~K OP converter temperature would correspond to a half-monolayer-thick film.

\section*{Acknowledgements}
This material is based upon work supported by the National Science Foundation under Grant No. PHY-1912425.
We gratefully acknowledge helpful conversations with Sunil Upadhyay and David Todd Anderson, and experimental assistance with the construction of our apparatus from Wade J. Cline, Carl D. Davidson, Jr., and Jake Holland. 
We thank the referees for their contributions to the paper during the review process.

\section*{Data availability}
The data that support the findings of this study are available from the corresponding author upon reasonable request.

\bibliography{High_purity_parahydrogen_2020.bib}

\end{document}